# FORECASTING TIME SERIES OF INHOMOGENEOUS POISSON PROCESSES WITH APPLICATION TO CALL CENTER WORKFORCE MANAGEMENT

By Haipeng Shen[1] and Jianhua Z. Huang[2]

*University of North Carolina at Chapel Hill and Texas A&M University*

We consider forecasting the latent rate profiles of a time series of inhomogeneous Poisson processes. The work is motivated by operations management of queueing systems, in particular, telephone call centers, where accurate forecasting of call arrival rates is a crucial primitive for efficient staffing of such centers. Our forecasting approach utilizes dimension reduction through a factor analysis of Poisson variables, followed by time series modeling of factor score series. Time series forecasts of factor scores are combined with factor loadings to yield forecasts of future Poisson rate profiles. Penalized Poisson regressions on factor loadings guided by time series forecasts of factor scores are used to generate dynamic within-process rate updating. Methods are also developed to obtain distributional forecasts. Our methods are illustrated using simulation and real data. The empirical results demonstrate how forecasting and dynamic updating of call arrival rates can affect the accuracy of call center staffing.

**1. Introduction.** Queueing models usually assume that arrivals to a queueing system follow an inhomogeneous Poisson process. For efficient management of such a system, accurate prediction of the underlying random rate function of the inhomogeneous Poisson process is of primary importance. For example, various staffing models need the forecasted rate as one essential input [Gans, Koole and Mandelbaum (2003)]. Both the latency and uncertainty of the rate makes such a prediction problem interesting and difficult. The current paper proposes a methodology for forecasting the arrival

Received July 2007; revised January 2008.

[1]Supported in part by NSF Grant DMS-06-06577.

[2]Supported in part by NSF Grant DMS-06-06580 and National Cancer Institute Grant CA57030.

rate functions of a time series of inhomogeneous Poisson processes. The proposal combines the Poisson process assumption with ideas of data-driven dimension reduction as well as time series forecasting.

Our primary motivation and application come from call centers, where customers contact their service providers through telephones and wait in *tele* queues. The daily call arrivals to a call center form a point process that can be modeled as an inhomogeneous Poisson process [Brown et al. (2005)]. Motivated by call center operations management, two forecasting problems will be addressed: to forecast future rate functions days or weeks ahead, and to update existing forecasts using additional information during a day. Throughout the paper, we will use call centers to facilitate our presentation; however, the developed methodology undoubtedly has a much wider application spectrum.

In this paper a daily call arrival process is modeled as a Poisson process with a stochastic intensity. In other words, a call arrival process is a Cox process [Cox (1955)], also known as a doubly stochastic Poisson process. However, instead of focusing on a single daily arrival process, we consider a time series of daily arrival processes. What makes our paper different from the literature of Cox processes is the time series dependency of the daily intensity functions. Such dependency makes it feasible to perform time series forecasts of the intensity functions using historical data.

Call centers have become a primary contact point between companies and their customers in modern business [Mandelbaum (2006), Aksin, Armony and Mehrotra (2007)]. The forecasting of future call arrival rates is critical for efficient agent staffing and scheduling, but the current practice is still in its infancy [Gans, Koole and Mandelbaum (2003)]. To close the gap, recently more research efforts have been devoted to call center forecasting. Among them, Weinberg, Brown and Stroud (2007) propose a model-based approach where a two-way multiplicative Bayesian model for inhomogeneous Poisson processes is used to predict the Poisson rate functions; while the data-driven approach of Shen and Huang (2008) combines dimension reduction through singular value decomposition (SVD) with time series forecasting, and provides a more robust solution to the problem that does not rely on the multiplicative model assumption. To the best of our knowledge, these are the only two papers that have dealt with the problem of dynamic intraday updating.

However, instead of modeling the arrival counts directly, both approaches model the square-root transformed counts, which are approximately normally distributed for Poisson counts [Brown et al. (2007)]. The data-driven approach in Shen and Huang (2008) does not make use of the Poisson assumption, and aims at forecasting future counts. To the extent that some stochastic model assumption is reasonable for the counts, such as inhomogeneous Poisson processes, incorporating it appropriately should improve the statistical efficiency of the method. Note that forecasting the latent arrival rates of the Poisson processes is of primary interest from the management point of view (see Section 4.1). Although the point forecasts for counts and rates are the same, their distributional forecasts are different. Furthermore, neither Weinberg, Brown and Stroud (2007) nor Shen and Huang (2008) has considered the managerial consequences of different forecasts of the rates of the Poisson processes.

The purpose of this paper is to develop a forecasting method that utilizes the widely used Poisson assumption in queueing models. More precisely, we consider a time series of inhomogeneous Poisson processes and use the historical data to forecast the future rate functions of the Poisson processes. In the call center application, for example, each Poisson process models the customer call arrivals during one day in a sequence of days. Since the rate function of each Poisson process is approximately constant during short time intervals (such as quarter hours in our data example in Section 4), the data for a given day can be aggregated into a vector of Poisson random variables with varying rates, counting the call arrivals during these intervals. The collection of such rates for a given day is referred as the *rate profile* and the corresponding counts form the *count profile*. The unobservable rate profiles form a vector time series, and our goal is to forecast future rate profiles using the observed historical count profiles.

The dimensionality of the vector time series of count profiles is usually high. Thus, we propose to first reduce dimension by building a factor model on the hidden rate profiles. The proposed dimension reduction technique can be viewed as an extension of singular value decomposition to Poisson data, or *Poisson SVD*. According to the factor model, every rate profile can be expressed as a linear combination of a few factors. Factor loadings and scores can be computed by fitting generalized linear models (GLM) [Dobson (2001)]. An alternating maximum likelihood algorithm is employed that fits marginal Poisson regression models by alternately fixing the factor loadings and scores. The algorithm is closely related to the alternating least squares algorithm for computing singular value decomposition [Gabriel and Zamir (1979)].

After the factor analysis, forecasting future rate profiles reduces to forecasting time series of factor scores. Univariate time series models can be built for each factor score series to produce time series forecasts of the rate profile. To achieve within-process dynamic rate updating, we propose to effectively combine the information contained in both the historical processes and the current process, through maximizing a penalized likelihood criterion. The criterion appropriately balances two measures, the goodness-of-fit of the factor model to the early segment of the current process and the deviation of the factor scores from their forecasts provided by the interday time series forecasting.

Our forecasting approach enjoys the benefit of both the model-driven approach of Weinberg, Brown and Stroud (2007) and the data-driven approach of Shen and Huang (2008). It directly models and forecasts the rate profiles of inhomogeneous Poisson processes. [Some earlier analysis is briefly documented in Shen, Huang and Lee (2007).] Making no rigid model assumptions on the rate profiles, our approach is robust, as shown in two simulation studies. In this paper we also illustrate how rate forecasting and dynamic updating benefits the staffing decision in a real call center application, which has not been considered before in the literature.

The data we analyze and use for building forecasting models have an interesting two-way structure: there are both day-to-day variation/dependence and within-day variation/dependence. This specific structure differentiates our work with the related literature on longitudinal data analysis and dynamic factor models, where only one-way structure is present. Longitudinal data analysis [Diggle et al. (2002)] concerns repeated measurements from many subjects. While there are typically within subject correlation, the measurements from different subjects are usually independent. Dynamic factor models are commonly used in econometrics for analyzing economic time series [Stock and Watson (2005)]. Since economic variables do not have a natural ordering as the one present in our within-day call volume profiles, intraday dynamic updating is not of concern in the dynamic factor models literature. Moreover, we are unaware of any work on dynamic factor models for Poisson count data.

The rest of the paper is structured as follows. Section 2 describes the factor model as well as the alternating maximum likelihood estimation algorithm for model estimation. Our forecasting approach is described in detail in Section 3. The proposed method is applied in Section 4 to the call center arrival data analyzed by Shen and Huang (2008). Some background on call center staffing is briefly discussed in Section 4.1. Section 4.2 fits the factor model to the real data. Two simulation studies are presented in Section 4.3.2 to illustrate the robustness of our method and its performance on forecasting hidden rate profiles. The effect of rate forecasting on staffing level is discussed in Section 4.3.3. In Section 4.4 several intraday rate updating methods are used to update the related staffing levels. We conclude in Section 5 with some discussion.

## 2. Dimension reduction of Poisson variables.

2.1. *A Poisson factor model.* Consider a time series of inhomogeneous Poisson processes that are observed over the same time interval. Following common practice, we assume that the arrival rate function of each Poisson process can be well approximated as being piecewise constant over short time intervals. Let $\mathbf{Y} = (y_{ij})$ be an $n \times m$ matrix that records the arrival counts from $n$ such processes with each process being aggregated into $m$ time intervals. Furthermore, we assume that $y_{ij}$ is a Poisson random variable with rate $\lambda_{ij}$. For notational purpose, we let $\boldsymbol{\Lambda} = (\lambda_{ij})$ denote the $n \times m$ hidden Poisson rate matrix. The $i$th row of $\boldsymbol{\Lambda}$, denoted as $\boldsymbol{\lambda}_{(i)}^T = (\lambda_{i1}, \ldots, \lambda_{im})$, is the *rate profile* of the $i$th process. Correspondingly, the $i$th row of $\mathbf{Y}$, denoted as $\mathbf{y}_{(i)}^T = (y_{i1}, \ldots, y_{im})$, is the *count profile* of the $i$th process. Note that the rows of $\boldsymbol{\Lambda}$ and $\mathbf{Y}$ are time ordered.

The rate profiles $\{\boldsymbol{\lambda}_{(i)}\}_{i=1}^n$ then form a vector-valued time series taking values in $\mathbb{R}^m$. It is of interest to forecast future rate profiles $\boldsymbol{\lambda}_{(n+h)}$ ($h > 0$). However, the difficulty is that the rate profiles are unobservable. The idea is to build a time series forecasting model on the corresponding count profiles $\{\mathbf{y}_{(i)}\}_{i=1}^n$, which can then be used to forecast $\boldsymbol{\lambda}_{(n+h)}$.

In practice, the dimensionality of the vector time series $\{\mathbf{y}_{(i)}\}$ is usually so large that it is infeasible to directly apply the classical vector autoregressive and moving average (VARMA) models [Reinsel (1997)]. For example, the call center application in Section 4 has $m = 68$, and the application of Weinberg, Brown and Stroud (2007) has $m = 169$. This calls for the necessity of dimension reduction. In addition, each $\mathbf{y}_{(i)}$ is a vector of Poisson random variables with a *positive* rate vector $\boldsymbol{\lambda}_{(i)}$. The Poisson nature of the data needs to be accounted for appropriately.

For dimension reduction of Poisson variables, we consider the following $K$-factor model,

$$(2.1) \qquad \begin{cases} \mathbf{y}_{(i)} \sim \text{Poisson}(\boldsymbol{\lambda}_{(i)}), & i = 1, \ldots, n, \\ g(\boldsymbol{\lambda}_{(i)}) = \beta_{i1} \mathbf{f}_1 + \cdots + \beta_{iK} \mathbf{f}_K \equiv \mathbf{F} \boldsymbol{\beta}_{(i)}, \end{cases}$$

where $\boldsymbol{\beta}_{(i)} = (\beta_{i1}, \ldots, \beta_{iK})^T$ is the $K$-vector of underlying factor scores for the $i$th rate profile, $\mathbf{F}_{m \times K} = (\mathbf{f}_1, \ldots, \mathbf{f}_K)$ contains the $K$ factor loading vectors in $\mathbb{R}^m$, and the transformation $g$ is a link function suitable for Poisson variables such as the logarithmic or square root function. Here and throughout the paper, application of the link function $g$ to a vector or matrix is understood as a componentwise operation. See McCullagh and Nelder (1989) and Dobson (2001) for more discussion on generalized linear models (GLM) and, in particular, specification of link functions.

Denote the factor matrix as $\mathbf{B}_{n \times K} = (\boldsymbol{\beta}_{(1)}^T, \ldots, \boldsymbol{\beta}_{(n)}^T)^T$. Then the factor model for the transformed rate profiles can be written in the following matrix form:

$$(2.2) \qquad \begin{cases} \mathbf{Y} \sim \text{Poisson}(\boldsymbol{\Lambda}), \\ g(\boldsymbol{\Lambda}) = \mathbf{B} \mathbf{F}^T. \end{cases}$$

The factor model summarizes the transformed rate profiles using $K$ common factors, denoted as $\boldsymbol{\beta}_1, \ldots, \boldsymbol{\beta}_K$, which are the columns of $\mathbf{B}$. Note that each $\boldsymbol{\beta}_k$ is a time series of factor scores. How to build time series models on these series for forecasting will be discussed in Section 3. In practice, one would assume $K$ is usually much smaller than $n$ or $m$, resulting in a considerable amount of dimension reduction. The above model (2.2) may appear similar to dynamic factor models (DFM) commonly used in econometrics [Stock and Watson (2005)]. However, there are two major differences. First, our loadings for the same process are time dependent and correlated, a useful feature for within-process updating discussed later in Section 3.2, while such dependence is missing for DFM because economic variables usually have no natural ordering. Second, our model deals with Poisson variables, which to the best of our knowledge have not been analyzed in the DFM literature.

As it stands, the factor model (2.2) is not identifiable. To achieve identifiability, one can require either $\mathbf{f}_k^T \mathbf{f}_{k'} = \delta_{kk'}$ or $\boldsymbol{\beta}_k^T \boldsymbol{\beta}_{k'} = \delta_{kk'}$, where $\delta_{kk'}$ is the Kronecker delta that equals 1 for $k = k'$ and 0 otherwise. See Section 2.2 for more details.

2.2. *An alternating maximum likelihood algorithm.* Assuming independence of data, one can estimate the factors $\mathbf{B}$ and the loadings $\mathbf{F}$ using maximum likelihood. However, since both $\mathbf{B}$ and $\mathbf{F}$ are unknown, direct maximization of the likelihood function is complicated. Below, we propose an iterative procedure that relies on alternating optimization over $\mathbf{B}$ and $\mathbf{F}$.

Fixing $\mathbf{F}$, the factor model (2.1) represents $n$ standard Poisson regression models with the count profiles $\mathbf{y}_{(i)}$ as responses and $\mathbf{F}$ as the common matrix of covariates. On the other hand, the same factor model can be written in terms of the columns of $\mathbf{Y}$ (denotes by $\{\mathbf{y}_j\}$) as

$$(2.3) \qquad \begin{cases} \mathbf{y}_j \sim \text{Poisson}(\boldsymbol{\lambda}_j), & j = 1, \ldots, m, \\ g(\boldsymbol{\lambda}_j) = \mathbf{B}\mathbf{f}_{(j)}. \end{cases}$$

Fixing $\mathbf{B}$, the model consists of $m$ Poisson regressions with $\mathbf{B}$ as the common regressor matrix.

The above discussion suggests the following iterative algorithm:

1. Initialize: Apply singular value decomposition (SVD) to $g(\mathbf{Y})$ to obtain the SVD components $\{\mathbf{U}, \mathbf{V}, \mathbf{S}\}$, and set $\mathbf{B}_{old} = [s_1\mathbf{u}_1, \ldots, s_K\mathbf{u}_K]$ and $\mathbf{F}_{old} = [\mathbf{v}_1, \ldots, \mathbf{v}_K]$, where $\mathbf{u}_k$ is the $k$th column of the left singular vector matrix $\mathbf{U}$, $\mathbf{v}_k$ is the $k$th column of the right singular vector matrix $\mathbf{V}$ and $s_k$ is the $k$th diagonal element of the ordered diagonal singular value matrix $\mathbf{S}$.
2. Update:
   (a) Fit the Poisson regression model (2.1) with $\mathbf{F}$ replaced by $\mathbf{F}_{old}$, and obtain an updated factor score matrix $\mathbf{B}_{\text{new}}$;
   (b) Fit the Poisson regression model (2.3) with $\mathbf{B}$ replaced by $\mathbf{B}_{\text{new}}$, and obtain an updated factor loading matrix $\mathbf{F}_{\text{new}}$;
   (c) Apply SVD to the product matrix $\mathbf{B}_{\text{new}}\mathbf{F}_{\text{new}}^T$, and denote the obtained SVD components by $\{\mathbf{U}_{\text{new}}, \mathbf{V}_{\text{new}}, \mathbf{S}_{\text{new}}\}$.

3. Repeat Step 2 with $\mathbf{B}_{old}$ and $\mathbf{F}_{old}$ replaced by the first $K$ columns of $\mathbf{U}_{\text{new}}\mathbf{S}_{\text{new}}$ and $\mathbf{V}_{\text{new}}$ respectively until convergence.

The algorithm is presented in such a way that the factor loadings are orthonormal in that $\mathbf{F}^T\mathbf{F} = I$. Alternatively, we can modify the algorithm to make the score vectors satisfy $\mathbf{B}^T\mathbf{B} = I$. To achieve that, we can set $\mathbf{B}$ to be the first $K$ columns of the left singular vector matrix $\mathbf{U}$, and $\mathbf{F}$ to be the corresponding columns of $\mathbf{SV}$. This alternative formulation turns out to be more useful for the penalized dynamic updating approach to be discussed in Section 3.2.

Our Poisson factor model extends singular value decomposition (SVD) to Poisson count data. The above alternating maximum likelihood algorithm is an extension of the alternating least squares algorithm for computing SVD [Gabriel and Zamir (1979)].

2.3. *Selection of the number of factors.* In practice, one needs to decide on $K$, the number of underlying factors. Since we are in a forecasting context, one possibility is to compare out-of-sample forecasting performance for different choices of $K$. Note that the forecasting performance measure can only be calculated using the counts, as the rates are not observed. Our experience suggests that increasing $K$ tends to improve forecasting performance, but the improvement becomes minimal beyond a particular value of $K$, denoted as $K^*$, as additional factors start to fit the noise in the data. This $K^*$ can be used as the "true" number of factors.

Alternatively, one can extend the idea of scree plots from principal components analysis (PCA) to the current context. In PCA, the eigenvalues of the sample covariance matrix are ordered decreasingly and plotted against their indices in a scree plot. An "elbow" in the scree plot, at which the slopes of lines joining the plotted points change from "steep" to "shallow," is usually used to locate the number of significant principal components (PCs). To extend scree plots to Poisson variables, note that the eigenvalues equal to the reductions of residual sum of squares in reconstruction of the data using PCs as more PCs are added.

In GLM, a commonly used goodness-of-fit measure is the so called *deviance* [Dobson (2001)]. Consider the $K$-factor model (2.1), and let $\widehat{\mathbf{B}}$ and $\widehat{\mathbf{F}}$ denote the maximum likelihood estimates of the parameters. Denote the corresponding maximized log-likelihood as

$$l(\mathbf{Y}; \mathbf{\Lambda} = \widehat{\mathbf{\Lambda}}) \qquad \text{with } \widehat{\mathbf{\Lambda}} = g^{-1}(\widehat{\mathbf{B}}\widehat{\mathbf{F}}^T).$$

In addition, consider the saturated model where the rate matrix $\mathbf{\Lambda}$ is estimated as the count matrix $\mathbf{Y}$, and denote the corresponding log-likelihood as $l(\mathbf{Y}; \mathbf{\Lambda} = \mathbf{Y})$. The deviance of the model (2.1) is then defined as

$$2\{l(\mathbf{Y}; \mathbf{\Lambda} = \mathbf{Y}) - l(\mathbf{Y}; \mathbf{\Lambda} = \widehat{\mathbf{\Lambda}})\}.$$

Similar to the residual sum of squares, the deviance decreases as $K$ increases. We propose to calculate the reduction of deviance of the model (2.1) for an increasing sequence of $K$, and plot it against $K$ in a *deviance reduction plot*. The number of necessary factors can then be suggested by an "elbow" in the plot. See Figure 1 for an illustration of this idea, which shows that the steep drop of the deviance slows down significantly beyond the elbow around four or five factors.

## 3. Forecasting.

3.1. *Forecasting future rate profile.* Given the historical count data $\mathbf{y}_{(i)}$, $i = 1, \ldots, n$, and assuming the latent Poisson rates satisfy (2.1), consider forecasting the future rate profile $\boldsymbol{\lambda}_{(n+h)}(h > 0)$. According to (2.1),

$$\boldsymbol{\lambda}_{(n+h)} = g^{-1}(\beta_{n+h,1}\mathbf{f}_1 + \cdots + \beta_{n+h,K}\mathbf{f}_K).$$

Since the factor loading vectors $\mathbf{f}_1, \ldots, \mathbf{f}_K$ can be obtained from historical data using the alternating maximum likelihood (AML) algorithm, forecasting the $m$-dimensional rate profile $\boldsymbol{\lambda}_{(n+h)}$ reduces to forecasting the $K$-dimensional vector of factor scores $\boldsymbol{\beta}_{(n+h)} = \{\beta_{n+h,1}, \ldots, \beta_{n+h,K}\}^T$. Suppose we can obtain such a forecast $\hat{\boldsymbol{\beta}}_{(n+h)}^{\mathrm{TS}}$, then a forecast of $\boldsymbol{\lambda}_{(n+h)}$ follows as

$$\hat{\boldsymbol{\lambda}}_{(n+h)}^{\mathrm{TS}} = g^{-1}(\hat{\beta}_{n+h,1}^{\mathrm{TS}}\mathbf{f}_1 + \cdots + \hat{\beta}_{n+h,K}^{\mathrm{TS}}\mathbf{f}_K),$$

where $\hat{\beta}_{n+h,k}^{\mathrm{TS}}$ is a time series forecast of $\beta_{n+h,k}$, $1 \leq k \leq K$. Because the count profile $\mathbf{y}_{(n+h)}$ has a Poisson distribution with rate $\boldsymbol{\lambda}_{(n+h)}$, the point forecast of $\mathbf{y}_{(n+h)}$ is the same as $\hat{\boldsymbol{\lambda}}_{(n+h)}^{\mathrm{TS}}$.

In addition to the factors, the AML algorithm also produces the matrix $\mathbf{B}$ of factor scores, whose row vectors $\boldsymbol{\beta}_{(1)}, \ldots, \boldsymbol{\beta}_{(n)}$ form a $K$-dimension time series. This vector time series is subject to time series modeling to generate desired forecast $\hat{\boldsymbol{\beta}}_{(n+h)}^{\mathrm{TS}}$. Note that the dimension $K$ of this vector time series is much smaller than the dimension $m$ of the original observation $y_{(i)}$. However, we propose to model the score time series for each factor separately using univariate time series models such as exponential smoothing, ARIMA models [Box, Jenkins and Reinsel (1997)] or state space models [Harvey (1990)]. An appropriate model can be decided on by analyzing historical data. The rational for this separate univariate time series modeling lies in the alternating estimation algorithm (Section 2.2). We observe that the score series $\boldsymbol{\beta}_k$ is orthogonal to score series $\boldsymbol{\beta}_{k'}'$ for $k \neq k'$. This lack of contemporaneous correlation suggests that the cross-correlations at nonzero lags are likely to be small. Hence, it suffices to forecast each $\boldsymbol{\beta}_k$ separately.

To generate interval and distributional forecasts, we modify the bootstrap approach described in Shen and Huang (2008). For a fixed $k$, we use the fitted time series model for the series $\boldsymbol{\beta}_k$ recursively, and bootstrap the model errors from the fitted model to generate a sample of $B$ forecasts $\{\hat{\beta}_{n+h,k}^{\text{TS},b}\}, 1 \leq b \leq B$. Then $B$ forecasts of $\boldsymbol{\lambda}_{(n+h)}$ can be obtained as

$$\widehat{\boldsymbol{\lambda}}_{(n+h)}^{\text{TS},b} = g^{-1}(\hat{\beta}_{n+h,1}^{\text{TS},b}\mathbf{f}_1 + \cdots + \hat{\beta}_{n+h,K}^{\text{TS},b}\mathbf{f}_K), \qquad b = 1, \ldots, B,$$

which provides a distributional forecast for $\boldsymbol{\lambda}_{(n+h)}$. To obtain a distributional forecast of the count profile, for any $b$, we randomly sample one count profile forecast $\widehat{\mathbf{y}}_{(n+h)}^{\text{TS},b}$ from Poisson distributions with rate $\widehat{\boldsymbol{\lambda}}_{(n+h)}^{\text{TS},b}$. The sample of $B$ forecasts of $\mathbf{y}_{(n+h)}$ then leads to its distributional forecast. Interval forecasts can be obtained easily using quantiles of the distributional forecasts.

### 3.2. Dynamic within-process rate updating.
In addition to the historical count profiles $\mathbf{y}_{(1)}, \ldots, \mathbf{y}_{(n)}$, sometimes one also observes the counts of the $(n+1)$th process during the first $m_0$ time intervals. It is of interest to use them to update $\widehat{\boldsymbol{\lambda}}_{(n+h)}^{\text{TS}}$, the time series forecast obtained previously. Such updating can reduce forecasting errors substantially [Weinberg, Brown and Stroud (2007), Shen and Huang (2008)]. For call centers, the ability to change agent schedules in response to the updating then yields operational benefits. For example, to adjust for a revised forecast, call center managers may send people home early or have them perform alternative tasks, or they may call in part-time or work-from-home agents.

Denote the new observations from the $(n+1)$th process collectively as $\mathbf{y}_{(n+1)}^e$ and the corresponding Poisson rate vector as $\boldsymbol{\lambda}_{(n+1)}^e$. Let $\mathbf{y}_{(n+1)}^l = (y_{n+1,m_0+1}, \ldots, y_{n+1,m})^T$ be the latter count profile for the $(n+1)$th process with the rate vector being $\boldsymbol{\lambda}_{(n+1)}^l$. For notational simplicity, we suppress the dependence of $\mathbf{y}_{(n+1)}^e/\boldsymbol{\lambda}_{(n+1)}^e$ and $\mathbf{y}_{(n+1)}^l/\boldsymbol{\lambda}_{(n+1)}^l$ on $m_0$ in the following discussion. There are two sets of information, $\{\mathbf{y}_{(1)}, \ldots, \mathbf{y}_{(n)}\}$ and $\mathbf{y}_{(n+1)}^e$. We are interested in making use of both to obtain an updated forecast for $\boldsymbol{\lambda}_{(n+1)}^l$.

Using historical data, the alternating maximum likelihood algorithm in Section 2.2 can be run to get estimates of the factor loading vectors $\mathbf{f}_1, \ldots, \mathbf{f}_K$ and the factor score series $\boldsymbol{\beta}_1, \ldots, \boldsymbol{\beta}_K$. Time series models are then built on $\boldsymbol{\beta}_1, \ldots, \boldsymbol{\beta}_K$ to generate forecasts $\hat{\beta}_{n+1,1}^{\text{TS}}, \ldots, \hat{\beta}_{n+1,K}^{\text{TS}}$. Therefore, just using the historical count profiles, the time series forecast of $\boldsymbol{\lambda}_{(n+1)}^l$ is given by

$$(3.1) \qquad \widehat{\boldsymbol{\lambda}}_{n+1}^{l,\text{TS}} = g^{-1}(\hat{\beta}_{n+1,1}^{\text{TS}}\mathbf{f}_1^l + \cdots + \hat{\beta}_{n+1,K}^{\text{TS}}\mathbf{f}_K^l),$$

where $\mathbf{f}_k^l$ is the latter segment of $\mathbf{f}_k$ after the initial $m_0$ time intervals. However, this forecast does not utilize any new information contained in $\mathbf{y}_{n+1}^e$.

The new information can be incorporated using a Poisson regression as we describe now. Suppose the factor model (2.1) holds for the rate profile

$\boldsymbol{\lambda}_{(n+1)}$, then we have the following expression,

$$(3.2) \qquad g(\lambda_{n+1,j}) = \beta_{n+1,1}f_{j1} + \cdots + \beta_{n+1,K}f_{jK}, \qquad j = 1, \ldots, m.$$

Let $\mathbf{F}^e$ be an $m_0 \times K$ matrix whose $(j,k)$th entry is $f_{jk}$, $1 \leq j \leq m_0$, $1 \leq k \leq K$, and let $\boldsymbol{\beta}_{(n+1)}$ denote the vector $(\beta_{n+1,1}, \ldots, \beta_{n+1,K})^T$ of factor scores. The following Poisson regression model can then be built on $\mathbf{y}^e_{(n+1)}$,

$$(3.3) \qquad \begin{cases} \mathbf{y}^e_{(n+1)} \sim \text{Poisson}(\boldsymbol{\lambda}^e_{(n+1)}), \\ g(\boldsymbol{\lambda}^e_{(n+1)}) = \mathbf{F}^e\boldsymbol{\beta}_{(n+1)}. \end{cases}$$

This suggests that we can forecast $\boldsymbol{\beta}_{(n+1)}$ by maximum likelihood (ML), solving the following optimization problem,

$$\min_{\boldsymbol{\beta}_{(n+1)}} \sum_{j=1}^{m_0} \{\lambda_{n+1,j} - y_{n+1,j} \log(\lambda_{n+1,j})\} \quad \text{subject to} \quad g(\boldsymbol{\lambda}^e_{(n+1)}) = \mathbf{F}^e\boldsymbol{\beta}_{(n+1)},$$

which minimizes the negative log-likelihood function. Note that the regressor $\mathbf{F}^e$ in the Poisson regression is obtained using historical data.

However, the above Poisson regression approach ignores the time series dependence among the rate profiles, present in the factor score series. To improve the ML forecast, we propose to combine it with the time series forecast of $\boldsymbol{\beta}_{(n+1)}$ using a penalized Poisson regression. Specifically, we minimize the following *penalized likelihood* criterion with respect to $\boldsymbol{\beta}_{(n+1)}$,

$$(3.4) \qquad \begin{aligned} \sum_{j=1}^{m_0} \{\lambda_{n+1,j} - y_{n+1,j} \log(\lambda_{n+1,j})\} + \omega \|\boldsymbol{\beta}_{(n+1)} - \hat{\boldsymbol{\beta}}^{\text{TS}}_{(n+1)}\|^2 \\ \text{subject to} \quad g(\boldsymbol{\lambda}^e_{(n+1)}) = \mathbf{F}^e\boldsymbol{\beta}_{(n+1)}, \end{aligned}$$

where $\hat{\boldsymbol{\beta}}^{\text{TS}}_{(n+1)}$ is a time series forecast based on the historical count profiles, and $\omega > 0$ is a penalty parameter. To simplify our procedure, only one penalty parameter is used. Thus, it is desirable for the $K$ time series $\{\boldsymbol{\beta}_k\}$ to be roughly on the same scale. This can be achieved by requiring them to be orthonormal in the alternating maximum likelihood algorithm (Section 2.2).

The penalized criterion (3.4) involves two terms: the first term measures the goodness-of-fit of the model to $\mathbf{y}^e_{(n+1)}$ in terms of likelihood, while the second term penalizes a large departure from the time series forecast. Its solution is a compromise between the two terms based on the size of $\omega$, the penalty parameter. In practice, $\omega$ can be selected based on the forecasting performance on a rolling hold-out sample [Shen and Huang (2008)].

Below in Section 3.3, we describe an iterative re-weighted least squares (IRLS) algorithm for minimizing (3.4) based on a quadratic approximation of the objective function. The minimizer then gives us the penalized maximum likelihood (PML) forecast of $\boldsymbol{\beta}_{(n+1)}$. The PML forecast of $\boldsymbol{\lambda}_{(n+1)}^l$ (and $\mathbf{y}_{(n+1)}^l$) is then given by

$$
(3.5) \qquad \hat{\boldsymbol{\lambda}}_{n+1}^{l,\text{PML}} = g^{-1}(\hat{\beta}_{n+1,1}^{\text{PML}}\mathbf{f}_1^l + \cdots + \hat{\beta}_{n+1,K}^{\text{PML}}\mathbf{f}_K^l).
$$

For distributional updates of $\boldsymbol{\lambda}_{(n+1)}^l$, we propose to use a bootstrap procedure similar to the one aforementioned in Section 3.1. First, by bootstrapping the errors from the time series forecasting models for $\boldsymbol{\beta}_{(n+1)}$, we obtain $B$ time series forecasts $\hat{\boldsymbol{\beta}}_{(n+1)}^{\text{TS},b}$, which in turn lead to $B$ PML forecasts $\hat{\boldsymbol{\beta}}_{(n+1)}^{\text{PML},b}$ by minimizing the criterion (3.4). The derivation of the $B$ PML forecasts can be sped up considerably via a one-step updating approximation as discussed at the end of Section 3.3. The $B$ PML updates of $\boldsymbol{\lambda}_{(n+1)}^l$ are constructed as follows:

$$
\hat{\lambda}_{n+1,j}^{\text{PML},b} = g^{-1}(\hat{\beta}_{n+1,1}^{\text{PML},b}f_{j1} + \cdots + \hat{\beta}_{n+1,K}^{\text{PML},b}f_{jK}),
$$
$$
j = m_0 + 1, \ldots, m; b = 1, \ldots, B.
$$

The interval and density forecasts of $\lambda_{n+1,j}$ are obtained using the empirical distribution of $\hat{\lambda}_{n+1,j}^{\text{PML},b}$, $b = 1, \ldots, B$. To obtain a distributional update of the count profile, for $b = 1, \ldots, B$, we randomly sample one forecast $\hat{y}_{n+1,j}^{\text{PML},b}$ of $y_{n+1,j}$ from the Poisson distribution with rate $\hat{\lambda}_{n+1,j}^{\text{PML},b}$. The sample of $B$ forecasts of $y_{n+1,j}$ then leads to its interval and density forecasts.

3.3. *An IRLS algorithm for penalized Poisson regression.* Suppose we have some initial estimate of $\boldsymbol{\beta}_{(n+1)}$ denoted $\boldsymbol{\beta}_{(n+1)}^0$, and the corresponding estimate of $\boldsymbol{\lambda}_{(n+1)}$ denoted as $\boldsymbol{\lambda}_{(n+1)}^0$. The minimizing criterion (3.4) is equivalent to

$$
\mathcal{C}(\boldsymbol{\beta}_{(n+1)}) \equiv \sum_{j=1}^{m_0}\{\lambda_{n+1,j} - y_{n+1,j}\log(\lambda_{n+1,j})\}
$$
$$
(3.6)
$$
$$
+ \omega \sum_{k=1}^{K}(\beta_{n+1,k} - \hat{\beta}_{n+1,k}^{\text{TS}})^2.
$$

Introducing $\boldsymbol{\lambda}_{(n+1)}^0$, the criterion (3.6) can be re-written as

$$
\mathcal{C}(\boldsymbol{\beta}_{(n+1)}) = \sum_{j=1}^{m_0}[(\lambda_{n+1,j} - \lambda_{n+1,j}^0)
$$
$$
- y_{n+1,j}\{\log(\lambda_{n+1,j}) - \log(\lambda_{n+1,j}^0)\}]
$$
$$
(3.7)
$$

$$+ \sum_{j=1}^{m_0} \{\lambda_{n+1,j}^0 - y_{n+1,j} \log(\lambda_{n+1,j}^0)\}$$

$$+ \omega \sum_{k=1}^{K} (\beta_{n+1,k} - \hat{\beta}_{n+1,k}^{\mathrm{TS}})^2,$$

where $\lambda_{n+1,j} = g^{-1}(\mathbf{f}_j^T \boldsymbol{\beta}_{(n+1)})$ and $\lambda_{n+1,j}^0 = g^{-1}(\mathbf{f}_j^{eT} \boldsymbol{\beta}_{(n+1)}^0)$. Dropping the sub/super-scripts, the summand in the first term of (3.7) is

$$(3.8) \qquad\qquad (\lambda - \lambda^0) - y\{\log(\lambda) - \log(\lambda^0)\},$$

where $\lambda = g^{-1}(\mathbf{f}^T \boldsymbol{\beta})$ and $\lambda^0 = g^{-1}(\mathbf{f}^T \boldsymbol{\beta}^0)$. It follows from a second-order Taylor expansion that

$$(\lambda - \lambda^0) - y\{\log(\lambda) - \log(\lambda^0)\} = w(y, \mathbf{f}^T \boldsymbol{\beta}^0)\{\mathbf{f}^T \boldsymbol{\beta} - y^*(y, \mathbf{f}^T \boldsymbol{\beta}^0)\}^2 + c(\boldsymbol{\beta}^0),$$

where $c(\boldsymbol{\beta}^0)$ is a constant given the initial estimate $\boldsymbol{\beta}^0$, while the specific expressions of the functions $w(\cdot, \cdot)$ and $y^*(\cdot, \cdot)$ depend on the link function $g$.

The exact expressions of $w(\cdot, \cdot)$ and $y^*(\cdot, \cdot)$ for several commonly used link functions are listed below. For the identity link,

$$w(y, \mathbf{f}^T \boldsymbol{\beta}^0) = \frac{y}{2(\mathbf{f}^T \boldsymbol{\beta}^0)^2}, \qquad y^*(y, \mathbf{f}^T \boldsymbol{\beta}^0) = \mathbf{f}^T \boldsymbol{\beta}^0 - \frac{(\mathbf{f}^T \boldsymbol{\beta}^0)^2 - y\mathbf{f}^T \boldsymbol{\beta}^0}{y};$$

for the logarithmic link,

$$w(y, \mathbf{f}^T \boldsymbol{\beta}^0) = \tfrac{1}{2}\mathbf{f}^T \boldsymbol{\beta}^0, \qquad y^*(y, \mathbf{f}^T \boldsymbol{\beta}^0) = \mathbf{f}^T \boldsymbol{\beta}^0 - (1 - ye^{-\mathbf{f}^T \boldsymbol{\beta}^0});$$

and for the square-root link,

$$w(y, \mathbf{f}^T \boldsymbol{\beta}^0) = 1 + \frac{y}{(\mathbf{f}^T \boldsymbol{\beta}^0)^2}, \qquad y^*(y, \mathbf{f}^T \boldsymbol{\beta}^0) = \mathbf{f}^T \boldsymbol{\beta}^0 - \frac{(\mathbf{f}^T \boldsymbol{\beta}^0)^3 - y\mathbf{f}^T \boldsymbol{\beta}^0}{(\mathbf{f}^T \boldsymbol{\beta}^0)^2 + y}.$$

Note that, for all three links, the induced response variable $y^*$ is a shifted version of $\mathbf{f}^T \boldsymbol{\beta}^0$.

Plugging back the sub/super-scripts, barring any additive constant independent of $\boldsymbol{\beta}_{(n+1)}$, the quadratic approximation of the minimizing criterion $\mathcal{C}(\boldsymbol{\beta}_{(n+1)})$ is equivalent to

$$\sum_{j=1}^{m_0} w(y_{n+1,j}, \mathbf{f}_j^{eT} \boldsymbol{\beta}_{(n+1)}^0)\{\mathbf{f}_j^{eT} \boldsymbol{\beta}_{(n+1)} - y^*(y_{n+1,j}, \mathbf{f}_j^{eT} \boldsymbol{\beta}_{(n+1)}^0)\}^2$$

$$+ \omega \sum_{k=1}^{K} (\beta_{n+1,k} - \hat{\beta}_{n+1,k}^{\mathrm{TS}})^2.$$

In matrix form, the above quadratic approximation can be written as

$$\mathcal{C}(\boldsymbol{\beta}_{(n+1)}) \approx (\mathbf{y}_{(n+1)}^* - \mathbf{F}^e \boldsymbol{\beta}_{(n+1)})^T \mathbf{W} (\mathbf{y}_{(n+1)}^* - \mathbf{F}^e \boldsymbol{\beta}_{(n+1)})$$
$$+ \omega (\boldsymbol{\beta}_{(n+1)} - \hat{\boldsymbol{\beta}}_{(n+1)}^{\mathrm{TS}})^T (\boldsymbol{\beta}_{(n+1)} - \hat{\boldsymbol{\beta}}_{(n+1)}^{\mathrm{TS}}),$$

where the induced response vector $\mathbf{y}_{(n+1)}^*$ has $y^*(y_{n+1,j}, \mathbf{f}_j^{eT} \boldsymbol{\beta}_{(n+1)}^0)$ as the $j$th entry, while the diagonal weighting matrix $\mathbf{W}$ has $w(y_{n+1,j}, \mathbf{f}_j^{eT} \boldsymbol{\beta}_{(n+1)}^0)$ as the $j$th diagonal entry. Hence, the minimizer of quadratic approximation of $\mathcal{C}(\boldsymbol{\beta}_{(n+1)})$ is

$$(3.9) \qquad \hat{\boldsymbol{\beta}}_{(n+1)} = (\mathbf{F}^{eT} \mathbf{W} \mathbf{F}^e + \omega \mathbf{I})^{-1} (\mathbf{F}^{eT} \mathbf{W} \mathbf{y}_{(n+1)}^* + \omega \hat{\boldsymbol{\beta}}_{(n+1)}^{\mathrm{TS}}).$$

The above derivation suggests the following iterative re-weighted least squares (IRLS) algorithm to solve the minimization problem (3.4):

1. Initialize: Fit the Poisson regression model (3.3) on $\mathbf{y}_{(n+1)}^e$ and use the maximum likelihood estimate of $\boldsymbol{\beta}_{(n+1)}$ as $\boldsymbol{\beta}_{(n+1)}^0$.

2. Update: Calculate the updated estimate $\hat{\boldsymbol{\beta}}_{(n+1)}$ according to (3.9) for a particular time series forecast $\hat{\boldsymbol{\beta}}_{(n+1)}^{\mathrm{TS}}$.

3. Repeat Step 2 with $\boldsymbol{\beta}_{(n+1)}^0$ replaced by newly updated $\hat{\boldsymbol{\beta}}_{(n+1)}$ until convergence.

When using the above IRLS algorithm to derive a distributional forecast for $\boldsymbol{\beta}_{(n+1)}$, the algorithm needs to be iterated until convergence for each bootstrapped time series forecast $\hat{\boldsymbol{\beta}}_{(n+1)}^{\mathrm{TS},b}$. In our implementation, we initialize the algorithm by setting $\boldsymbol{\beta}_{(n+1)}^0$ to be the penalized point forecast $\boldsymbol{\beta}_{(n+1)}^{\mathrm{PML}}$, and iterate the algorithm one step toward the bootstrapped forecast $\hat{\boldsymbol{\beta}}_{(n+1)}^{\mathrm{TS},b}$. This simplification helps reduce the computational effort and works well in our examples.

**4. Application to call center data.** In this section we illustrate the proposed methods using the call center data analyzed in Shen and Huang (2008). For efficient staffing of the call center, we are interested in forecasting future daily arrival rate profiles (Section 4.3), and dynamically updating existing forecasts using new information within a day (Section 4.4). Shen and Huang (2008) focused on forecasting future call volumes instead of rates. In addition to extending their work to deal with Poisson rate forecasting, we go one step beyond comparing various rate forecasts—the forecasted rates are used to calculate the required staffing level, that is, the number of agents necessary to staff the call center in order to achieve some service level constraint (Sections 4.3.3 and 4.4). The effect of rate forecasting and updating on determining the staffing level has not been considered before.

The call center data record the incoming call volumes to a northeastern US bank call center during every quarter hour within the normal business hour (7:00AM–midnight). The data cover 210 weekdays between January 6th and October 24th, 2003. Among them, ten abnormal days are excluded from the analysis, including six holidays where the call volumes are very low and four days where the data are missing [Shen and Huang (2008)]. Empirical research has recently suggested that it is appropriate to model the arrival process of customer calls to a call center as an inhomogeneous Poisson process [Brown et al. (2005)].

This section is organized as follows. We first provide some background on call center agent staffing in Section 4.1 to facilitate our later calculation of staffing levels. We fit the factor model (2.1) in Section 4.2 and develop time series models for the factor score series. In Section 4.3 simulation studies are first performed to investigate the rate forecasting performance of our methods. The effect of interday rate forecasting on staffing level is then illustrated using the real data. Various intraday updating methods are then compared in Section 4.4 in terms of the related staffing level forecasts.

4.1. *Background on call center agent staffing.* Service systems such as call centers are usually analyzed as queueing systems. Queueing theoretic models, such as Erlang-C (or $M/M/N$ queue), are used to balance agent utilization and quality of service (QoS) according to some pre-specified QoS measure [Gans, Koole and Mandelbaum (2003)]. The capacity planning and workforce scheduling process in such systems usually takes place in three steps. First, historical arrival data are used to generate forecasts of arrival rates in each time period over some planning horizon. Second, queueing models are used to determine staffing levels for each of the time periods. For example, one simplistic yet previously commonly-used staffing strategy is the 80-20 rule, which finds the minimum number of agents to staff a system such that "80% of the customers will be served while their waiting (or delay) time is less than 20 seconds." Finally, the staffing level obtained in the second step is used as inputs to some mathematical programming to generate a set of agent schedules.

Below we focus on the second step of deciding the staffing level, and illustrate how improved rate forecasting, especially dynamic intraday updating, can improve the staffing accuracy. Suppose the arrival rate during a time period is $\lambda$, and an agent can handle $\mu$ calls per unit of time. Then, the *offered load* is defined as $R = \lambda/\mu$. Intuitively, this gives the minimum number of agents that a call center needs in order for the system to be stable. Otherwise, more calls will arrive than the agents can process, and the queue will eventually explode.

One well-known heavy traffic queueing-theoretic result [Halfin and Whitt (1981)] is that for highly utilized moderate to large systems such as the call center we analyze here, the required staffing level $N$ is approximately equal to

$$(4.1) \qquad N = R + \theta\sqrt{R},$$

where $\theta > 0$ is a service level parameter that is decided by the steady-state probability of calls being delayed, $\alpha$. To be exact,

$$(4.2) \qquad \alpha = \left\{ 1 + \frac{\theta\Phi(\theta)}{\phi(\theta)} \right\}^{-1},$$

where $\Phi$ and $\phi$ are the standard normal distribution and density functions. For example, $\theta = 1$ corresponds to 22% of the calls being delayed in steady state.

This simple staffing rule is known as the *square-root safety staffing principle*, where $\theta\sqrt{R}$ is the safety staffing needed above the minimum requirement $R$ in order to achieve some desired service level. The reason that the safety staffing is in the order of the square-root of the offered load is nontrivial. Recently, Garnett, Mandelbaum and Reiman (2002) extend the above result to queueing systems with abandonment by allowing the service level parameter $\theta$ to be negative. In call centers, some impatient customers end up abandoning the tele-queue before getting served. Square-root safety staffing is awfully simple. However, in order for it to work well in practice, one relies on accurate prediction of future offered load $R$, in particular, the arrival rate $\lambda$, which is the focus of the current research.

4.2. *Model building.* We considered $K = 1, \ldots, 10$ and fit the corresponding factor model (2.1) to the call center data. The square-root link function was employed. The left panel of Figure 1 gives the deviance reduction plot, which suggests that five factors are sufficient as the reduction of the deviance becomes flat afterward. The right panel plots the first five factor loading vectors, which are normalized to achieve comparable scales. The first loading vector roughly summarizes the average daily rate profile. The additional loading vectors capture various contrasts among different time periods within a day. The findings are consistent with those in Shen and Huang (2008) except theirs are about the count (instead of rate) profiles.

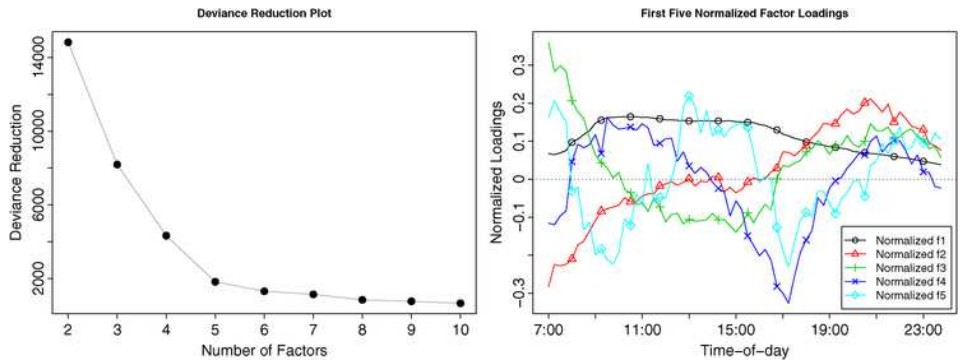

FIG. 1.  *The deviance reduction plot, suggesting five significant factors whose loadings are plotted.*

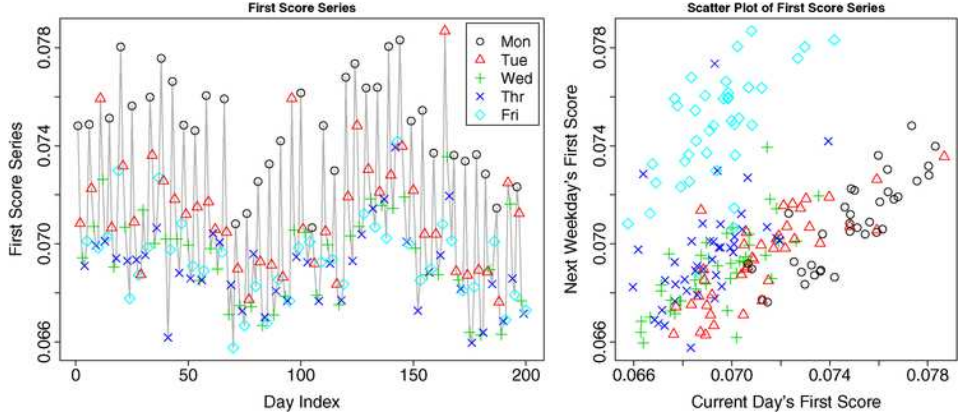

Fig. 2.  *Time series plot of the first score series and its lag-one scatter plot, suggesting an AR(1) time series model with a day-of-the-week dependent slope.*

To perform interday forecasting and intraday updating, we first need to develop a forecasting model for each score time series. The left panel of Figure 2 plots the first score series $\boldsymbol{\beta}_1$. Different symbols indicate different weekdays, revealing a strong day-of-the-week effect. The right panel shows that the factor scores of two consecutive weekdays are linearly related, conditioning on day-of-the-week. This motivated us to consider the following varying-coefficient AR(1) model,

$$\beta_{i1} = a_1(d_{i-1}) + b_1\beta_{i-1,1} + \varepsilon_{i1},$$

where $d_{i-1}$ denotes the day-of-the-week of day $i-1$, and the varying intercept $a_1$ depends on $d_{i-1}$. A more complicated model was also considered where the slope is allowed to depend on day-of-the-week. However, an $F$-test for the nested models returns a $p$-value of 0.7724, which suggests that the larger model does not provide any significant improvement.

Further exploratory data analysis suggested similar varying intercept AR(1) models for the other score series. Hence, we opt to use such models to forecast the future scores, which are denoted as

$$(4.3) \qquad \beta_{ik} = a_k(d_{i-1}) + b_k\beta_{i-1,k} + \varepsilon_{ik}, \qquad i = 2, \ldots, n, k = 1, \ldots, 5.$$

4.3. *One-day-ahead rate forecasting.* We first describe in Section 4.3.1 two useful parametric models motivated by the call center data. The models are used in Section 4.3.2 to generate data in simulation studies to test our interday forecasting methods. By considering the effect of forecasting on staffing level, our forecasting methods will be further tested in Section 4.3.3 using the real data. For our methods, we consider dimension reduction by the factor model (2.1) with $K = 1, \dots, 5$ underlying factors respectively, followed by time series modeling with (4.3) on each series of factor scores. We shall denote our methods as TS1 to TS5 depending on the number of factors used.

4.3.1. *Two parametric models.* We consider two parametric models, where the data $y_{ij} \sim \text{Poisson}(\lambda_{ij})$, $i = 1, \dots, n$, $j = 1, \dots, m$, and the underlying Poisson rate is assumed to have a two-way structure characterizing respectively the day-by-day variation and time-of-day variation. These two kinds of variations are typical for call center data [Brown et al. (2005), Weinberg, Brown and Stroud (2007)]. By taking into count the Poisson nature of the observations, the two models extend the corresponding models used in Shen and Huang (2008), which instead assume the square-root transformed counts as Gaussian observations.

*A multiplicative model.* The multiplicative (MUL) model assumes that the Poisson rates satisfy

$$(4.4) \quad \begin{cases} \sqrt{\lambda_{ij}} = \alpha_i \gamma_{d_i,j}, & d_i = 1, 2, 3, 4, 5, \\ \alpha_i - a_{d_i} = b(\alpha_{i-1} - a_{d_{i-1}}) + \eta_i, & \eta_i \sim \mathcal{N}(0, \phi^2), \\ \sum_j \gamma_{d_i,j} = 1. \end{cases}$$

Under this model, the square-root of the rate has a two-way multiplicative structure: the day-to-day variation ($\alpha_i$) is auto-regressive of order one with mean depending on day-of-the-week ($d_i$); the time-of-day variation ($\gamma_{d_i,j}$) also depends on day-of-the-week. This model can be viewed as a non-Bayesian version of the Bayesian model proposed by Weinberg, Brown and Stroud (2007), except that they also require the intraday variation $\gamma_{d_i,j}$ to be smooth.

*An additive model.* The additive (ADD) model assumes that the Poisson rates satisfy

$$(4.5) \quad \begin{cases} \sqrt{\lambda_{ij}} = \mu + \alpha_i + \beta_j + \gamma_{d_i,j}, & d_i = 1, 2, 3, 4, 5, \\ \alpha_i - a_{d_i} = b(\alpha_{i-1} - a_{d_{i-1}}) + \eta_i, & \eta_i \sim \mathcal{N}(0, \phi^2), \\ \sum_i \alpha_i = \sum_j \beta_j = \sum_i \gamma_{d_i,j} = \sum_j \gamma_{d_i,j} = \sum_{ij} \gamma_{d_i,j} = 0. \end{cases}$$

Under this model, the square-root-transformed rate has a two-way additive structure: the day-to-day variation follows an AR(1) process adjusting for day-of-the-week ($a_{d_i}$) effect; it also has an interaction with the time-of-day variation ($\gamma_{d_i,j}$).

As mentioned earlier, when $y_{ij} \sim \text{Poisson}(\lambda_{ij})$, $x_{ij} = \sqrt{y_{ij} + 1/4}$ is approximately $\mathcal{N}(\sqrt{\lambda_{ij}}, 1/4)$. Hence, the estimation and forecasting methods described in Sections A.3 and A.4 of Shen and Huang (2008) are applicable for the above two models.

Both Weinberg, Brown and Stroud (2007) and Shen and Huang (2008) considered an industry standard as their benchmark, which uses historical averages of the same day-of-the-week as forecasts. They both showed that the approach works poorly; hence, we choose not to consider it in our studies.

4.3.2. *Simulation studies of rate forecasting performance.* Because the Poisson rate profiles are unobservable in practice, in order to evaluate the rate forecasting performance of our methods, we carry out two simulation studies, using the multiplicative model (4.4) and the additive model (4.5) respectively to generate data. The model parameters are decided on by fitting the models to the real data described at the beginning of Section 4 and rounding the parameter estimates properly. We generate 100 data sets from each model for use in our simulation studies. The simulation results (detailed below) show that, without a specific parametric model assumption, our approach can always perform almost as well as the method using the true data generating model. As a comparison, a method based on a specific model assumption performs worse if the model assumption is wrong.

Seven forecasting methods are considered in our simulation studies, including forecasting assuming the multiplicative model (4.4), forecasting assuming the additive model (4.5), and our methods TS1 to TS5. For all methods, the true square-root link function is used. For each data set, a rolling out-of-sample forecast exercise is performed with the last 50 days as the forecasting set; and for each day in the forecasting set, the preceding 150 days are used to derive the forecast. The parameters involved in the methods are re-estimated for each day in the forecasting set.

To compare the performance of forecasting the underlying rate, two performance measures are calculated, the root mean squared error (RMSE) and mean relative error (MRE). For day $i$ in each data set, we define

$$\text{RMSE}_i = \sqrt{\frac{1}{m} \sum_{j=1}^{m} (\hat{\lambda}_{ij} - \lambda_{ij})^2} \quad \text{and} \quad \text{MRE}_i = \frac{100}{m} \sum_{j=1}^{m} \frac{|\hat{\lambda}_{ij} - \lambda_{ij}|}{\lambda_{ij}},$$

where $\hat{\lambda}_{ij}$ is the forecast for $\lambda_{ij}$.

The Mean RMSE and Mean MRE (%) of the forecasted rates are then calculated for each simulated data set. For each forecasting method, the two panels of Figure 3 plot the empirical cumulative distribution functions (CDF) of the Mean RMSE calculated on the 100 data sets simulated using the multiplicative and additive models respectively. The true model is indicated in the legend using an asterisk. When the data are generated using the multiplicative model, the additive model is a misspecified model and vice versa. In both cases, our TS methods do not use the model assumptions on the form of the underlying Poisson rate. The improvement of TS5 over TS4 is minimal; hence, the CDF for TS5 is omitted in both panels.

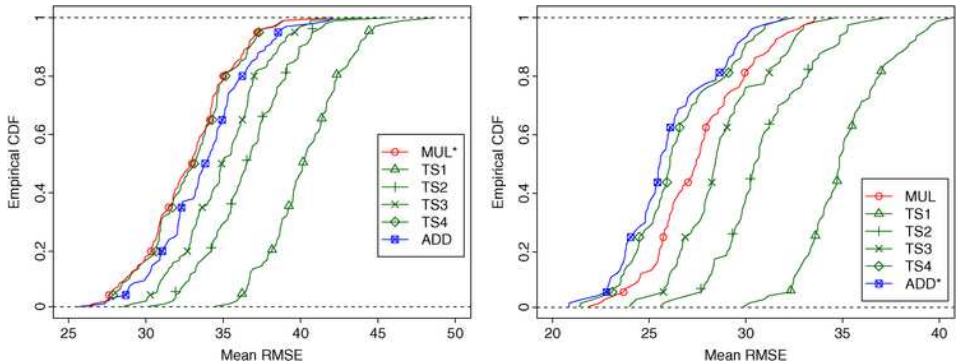

Fig. 3. *Comparison of empirical CDF of Mean RMSE for rate forecasting. The true model is highlighted in the legend using an asterisk. TS4 performs comparably to the true model, while the wrong model performs worse.*

Several observations can be made from the plots. For the TS methods, when increasing the number of factors $K$, the performance measure gets stochastically smaller; TS4 is competitively close to the true model; when $K$ increases beyond five, the forecasting improvement remains minimal. This suggests that four underlying factors are sufficient to approximate the true models. In addition, the wrong parametric model results in stochastically inferior forecasts than TS4. The uniformly good performance of our method shows its robustness against model assumptions, which is important in practice where the information of the true underlying model is hardly ever available. The results for Mean MRE (%) are similar and are not shown.

4.3.3. *The effect of rate forecasting on staffing level.* Below we apply the seven forecasting methods considered previously in Section 4.3.2 to the call center data, and illustrate how the rate forecasting affects staffing level. This important fundamental question has not been investigated previously in the literature. To calculate the staffing level, we apply the square-root safety staffing rule discussed in Section 4.1.

We assume an average service time of 5 minutes per call, which corresponds to an agent service rate of 12 calls per hour, or 3 calls per 15-minute interval. This figure is chosen subjectively for illustration purposes based on the empirical analysis in Brown et al. (2005). Suppose the arrival rate is

*Summary statistics (mean, median, lower quartile Q1, upper quartile Q3) of RMSE and MRE of the forecasted staffing level in a rolling forecast exercise. The forecasting set contains 50 days. MUL, TS4 and TS5 perform comparable*

| | **RMSE** | | | | **MRE (%)** | | | |
|------|-------|--------|-------|-------|-------|--------|-------|-------|
| | **Q1** | **Median** | **Mean** | **Q3** | **Q1** | **Median** | **Mean** | **Q3** |
| ADD | 14.28 | 18.03 | 20.31 | 21.80 | 4.63 | 5.40 | 6.29 | 6.68 |
| MUL | 14.27 | 16.65 | 19.69 | 20.38 | 4.46 | **5.06** | **5.91** | **6.30** |
| TS1 | 14.91 | 18.55 | 20.63 | 21.79 | 4.96 | 6.28 | 7.07 | 8.31 |
| TS2 | **13.54** | 17.51 | 19.88 | 20.89 | 4.73 | 5.41 | 6.32 | 6.88 |
| TS3 | 14.74 | 17.80 | 19.77 | **19.90** | 4.63 | 5.27 | 6.12 | 6.62 |
| TS4 | 14.32 | **16.55** | 19.50 | 20.27 | 4.45 | **5.06** | 6.00 | 6.42 |
| TS5 | 13.98 | **16.55** | **19.46** | 20.57 | **4.44** | **5.06** | 5.98 | 6.35 |

forecasted to be $\lambda_j$ for the $j$th interval. Then the offered load is $R_j = \lambda_j/3$. To figure out the additional safety staffing, we assume that the call center expects a 22% delay probability in steady state, which then determines the safety staffing factor $\theta$ to be 1 according to (4.2). Hence, it follows that the required staffing level for interval $j$ is

$$\lambda_j/3 + \sqrt{\lambda_j/3}.$$

We perform, on our data set, the same rolling one-day-ahead forecasting exercise as the one described in the simulation studies of Section 4.3.2. As the benchmark, we use the "oracle" staffing level decided by assuming the actual call volumes as the rate forecasts. Note that we wouldn't observe the actual volumes when actually planning the staffing in practice. We then treat the "oracle" staffing level as the truth when calculating performance measures of the forecasted staffing level decided by the various rate forecasts.

Table 1 compares summary statistics of the RMSE and MRE of the staffing level forecasts from the seven methods. For the TS methods, the results from using the square-root link function are presented. In general, the forecasting accuracy improves as one increases $K$, the number of factors. TS4 and TS5 give comparable results. On the ground of parsimony, TS4 may be preferred to TS5. The ADD model performs similar to TS2, while MUL similar to TS4/TS5.

In practice, out-of-sample forecasting performance can be used to pick the link function. Our choice of the square-root link in Table 1 is primarily for comparison purpose, because both ADD and MUL models use the square-root link. In addition, we also tried the identity link and the logarithmic link for our methods in this example, which generated inferior results than the square-root link.

4.4. *Dynamic intraday updating of rate and staffing level.* The call center data are used to illustrate the benefit of dynamic updating. We focus on the 10:00AM updating and the 12:00PM updating, which means that intraday dynamic updating is performed at 10:00AM and 12:00PM, respectively. The benchmark is the TS4 method which performs no updating. We shall use the square-root link and $K = 4$ based on the forecasting results in Section 4.3.3. Correspondingly, we term our penalized maximum likelihood intraday updating approach as PML4.

For comparison purpose, we also consider two alternative intraday updating approaches that combine the MUL/ADD forecasts with historical proportions (HP). Suppose the MUL point forecast for the rate profile of day $n + 1$ is $\hat{\boldsymbol{\lambda}}_{(n+1)}^{\mathrm{MUL}}$, which also gives the point forecast for the count profile $\mathbf{y}_{(n+1)}$. For an updating point $m_0$:

- calculate the ratio $R$ between *the total number of calls arrived* and *the cumulative forecasted rate* up to the time period $m_0$,

$$R = \frac{\sum_{j=1}^{m_0} y_{n+1,j}}{\sum_{j=1}^{m_0} \hat{\lambda}_{n+1,j}^{\mathrm{MUL}}};$$

- update the MUL forecasts for the remainder of the day as

$$\hat{\lambda}_{n+1,j}^{\mathrm{HPM}} = R\hat{\lambda}_{n+1,j}^{\mathrm{MUL}}, \qquad j = m_0 + 1, \ldots, m.$$

We refer to the above updating method as HPM. One can replace the MUL forecast with the ADD forecast in the above algorithm, which leads to the HPA updating. The HPM and HPA updating methods given here are extensions of similar methods for updating counts described in Section 4.2.2 of Shen and Huang (2008).

Below we apply the aforementioned updating methods to the call center data, and illustrate the effect of dynamic intraday updating on staffing call centers. The same rolling forecasting exercise as in Section 4.3.3 is performed, and we again use the square-root staffing rule to determine staffing level.

For PML4, the value of the penalty parameter $\omega$ needs to be decided at each updating point. To this end, we consider a set of candidate values, and choose the one that minimizes some forecasting performance measure based on the call volumes (rather than rates), such as RMSE. Specifically, we use the beginning 150 days in our data set as the training set, and perform an out-of-sample rolling forecast exercise on this training set. Note that this training set does not overlap with the forecasting set we initially hold out for out-of-sample forecast evaluation. For the purpose of selecting $\omega$, the last one third (i.e., 50 days) of the training set is used as a hold-out sample. For each day in this sample, its preceding 100 days are used to fit the Poisson factor model and to generate the PML4 updating for each given $\omega$. Compute some

TABLE 2

*Summary statistics (mean, median, lower quartile Q1 and upper quartile Q3) of RMSE of the updated staffing level for the 10:00AM and 12:00PM updatings. PML4 outperforms the other methods*

|      | **10:00AM updating** | | | | **12:00PM updating** | | | |
|------|-------|--------|-------|-------|-------|--------|-------|-------|
|      | **Q1** | **Median** | **Mean** | **Q3** | **Q1** | **Median** | **Mean** | **Q3** |
| TS4  | 12.42 | 14.29 | 18.03 | 18.82 | 12.42 | 14.29 | 18.03 | 18.82 |
| HPA  | 12.80 | 17.04 | 19.33 | 21.46 | 12.20 | 13.23 | 16.67 | 16.33 |
| HPM  | 12.71 | 15.49 | 19.02 | 21.03 | 11.98 | 13.04 | 16.17 | 16.02 |
| PML4 | 12.62 | 14.81 | 17.19 | 18.10 | **11.46** | **12.87** | **15.41** | **14.54** |

performance measure (e.g., RMSE) for every day in the hold-out sample and take the average. The $\omega$ that minimizes this average performance measure will be used for all days in the forecasting set. In this study, we consider $\omega$ from $\{0, 10, \ldots, 10^9\}$, and $\omega = 10^3$ is chosen for both updatings.

Table 2 presents summary statistics of the RMSE of the forecasted staffing levels from TS4, HPA, HPM and PML4. The averages are calculated over the 50 days in the forecasting set. For a fair comparison, only data after 12:00PM are used when calculating the RMSE. The superior performance of PML4 over the other methods is quite clear, which improves over TS4 by 14.5% on average. We also observe that, for every intraday updating method, updating later always improves the forecasting accuracy. In addition, intraday updating reduces the prediction interval width as illustrated in the right panel of Figure 4.

Figure 4 provides a graphical illustration of the forecasted staffing levels for September 2nd, 2003. The left panel shows the required number of agents for every 15-minute interval between 10:00AM and 17:00PM, when the majority of the calls get connected to the center. We observe that the 10:00AM updating results in a dramatic upward shift in the related staffing level, which is very close to the oracle staffing level. On the other hand, the TS4 forecast leads to uniformly under-staffing throughout the day. What happens is that September 2nd corresponds to the day after Labor Day. Consequently, the call center experienced an unusably high call volume that day, which was not accounted for by the TS4 method. The HP updating methods can somehow adjust for the upward shift; however, their performance is worse than the PML4 method.

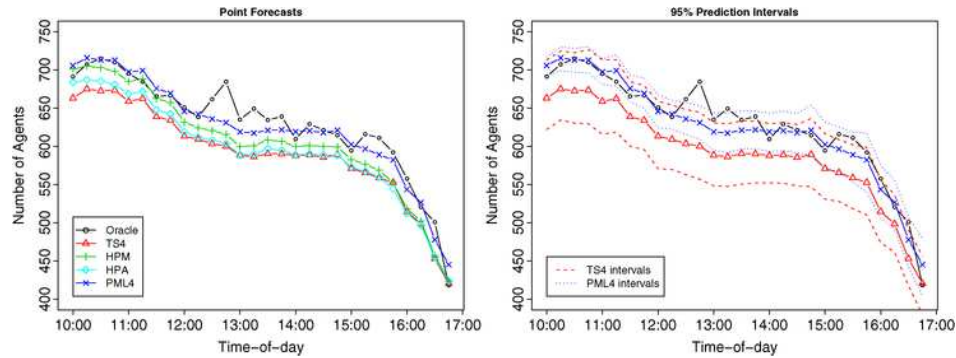

FIG. 4. *Comparison between the required number of agents per 15-minute interval for September 2nd based on the forecasted rates from various methods. The PML4 10:00AM updating gives the most accurate staffing.*

In the right panel of Figure 4, we superimpose the 95% prediction intervals for the staffing level resulting from the corresponding intervals for the forecasted arrival rates. Only PML4 and TS4 are compared. The number of bootstrapped forecasts is chosen to be $B = 1000$. We observe that the intervals from the intraday updating (PML4) almost always cover the "true" staffing level, while the intervals without the updating miss for more than half of the time periods. In addition, intraday updating results in uniformly narrower prediction intervals, which suggests that the prediction is more precise. The average interval width for TS4 and PML4 is 84.86 and 49.65, respectively; hence, a 41.5% average reduction results from the intraday updating. The managerial benefit is that the call center manager can work with a much narrower range of staffing level in the presence of arrival rate uncertainty.

**5. Conclusion and discussion.** In this paper we develop methods for forecasting and dynamic updating of the latent and uncertain rate profiles of a time series of inhomogeneous Poisson processes. Our new approach extends the model-driven approach of Weinberg, Brown and Stroud (2007) and the data-driven approach of Shen and Huang (2008). The latter does not explicitly consider forecasting the latent Poisson rates. Instead, the authors consider forecasting of call volumes that can be regarded as a noisy version of the rates. We propose a Poisson factor model that combines the distribution assumption with data-driven dimension reduction. An iterative algorithm is proposed for parameter estimation. Besides showing the competitiveness and robustness of our forecasting methods, we also illustrate how intraday updating can improve the accuracy of call center staffing, which is the first in the relevant literature and provides some important managerial insights.

In terms of model fitting, our approach involves two stages: first fit a Poisson factor model for dimension reduction, then fit time series models on the factor score series. If the time series models could be specified together with the factor model prior to data analysis, a full likelihood approach could be statistically more efficient. However, the time series models are usually not easy to specify beforehand, so our two stage procedure is natural for model building. Moreover, numerical calculation for the full likelihood approach is much more complicated than our procedure, although it is conceptually simple. In our call center application, fast computation is of great concern, especially when performing dynamic within-day updating. Nevertheless, it is still interesting to explore the full likelihood approach. Whether statistical efficiency of using full likelihood translates into benefits in accurate forecasting deserves further research.

**Acknowledgments.** The authors want to extend grateful thanks to the Editor, the Associate Editor and two reviewers, whose comments have greatly improved the scope and presentation of the paper.

## REFERENCES

Aksin, Z., Armony, M. and Mehrotra, V. (2007). The modern call-center: A multidisciplinary perspective on operations management research. *Production and Operations Management* **16** 665–688.

Box, G. E. P., Jenkins, G. M. and Reinsel, G. C. (1994). *Time Series Analysis: Forecasting and Control*, 3rd ed. Prentice Hall, Englewood Cliffs, New Jersey. MR1312604

Brown, L. D., Cai, T., Zhang, R., Zhao, L. and Zhou, H. (2007). The root-unroot algorithm for density estimation as implemented via wavelet block thresholding. Technical report.

Brown, L. D., Gans, N., Mandelbaum, A., Sakov, A., Shen, H., Zeltyn, S. and Zhao, L. (2005). Statistical analysis of a telephone call center: A queueing-science perspective. *J. Amer. Statist. Assoc.* **100** 36–50. MR2166068

Cox, D. R. (1955). Some statistical methods connected with series of events. *J. Roy. Statist. Soc. Ser. B* **17** 129–164. MR0092301

Diggle, P., Heagerty, P., Liang, K. Y. and Zeger, S. (2002). *Analysis of Longitudinal Data*, 2nd ed. Oxford Univ. Press, New York. MR2049007

Dobson, A. J. (2001). *An Introduction to Generalized Linear Models*, 2nd ed. Chapman and Hall/CRC, London. MR1070981

Gabriel, K. R. and Zamir, S. (1979). Lower rank approximation of matrices by least squares with any choice of weights. *Technometrics* **21** 489–498.

Gans, N., Koole, G. M. and Mandelbaum, A. (2003). Telephone call centers: Tutorial, review, and research prospects. *Manufacturing & Service Operations Management* **5** 79–141.

Garnett, O., Mandelbaum, A. and Reiman, M. (2002). Designing a call center with impatient customers. *Manufacturing & Service Operations Management* **4** 208–227.

Halfin, S. and Whitt, W. (1981). Heavy-traffic limits for queues with many exponential servers. *Operations Research* **29** 567–588. MR0629195

Harvey, A. C. (1990). *Forecasting, Structural Time Series Models and the Kalman Filter*. Cambridge Univ. Press, New York. MR1085719

Mandelbaum, A. (2006). Call centers. Research bibliography with abstracts. Technical report, Technion, Israel.

McCullagh, P. and Nelder, J. A. (1989). *Generalized Linear Models*, 2nd ed. Chapman and Hall, London. MR0727836

Reinsel, G. C. (1997). *Elements of Multivariate Time Series Analysis*, 2nd ed. Springer, New York. MR1451875

Shen, H. and Huang, J. Z. (2008). Interday forecasting and intraday updating of call center arrivals. *Manufacturing & Service Operations Management*. To appear.

SHEN, H., HUANG, J. Z. and LEE, C. (2007). Forecasting and dynamic updating of uncertain arrival rates to a call center. *Proceedings of the IEEE/INFORMS International Conference on Service Operations and Logistics, and Informatics* 50–55.

STOCK, J. H. and WATSON, M. W. (2005). Implications of dynamic factor models for VAR analysis. NBER Working Paper No. 11467.

WEINBERG, J., BROWN, L. D. and STROUD, J. R. (2007). Bayesian forecasting of an inhomogeneous Poisson process with applications to call center data. *J. Amer. Statist. Assoc.* **102** 1185–1199.

DEPARTMENT OF STATISTICS AND OPERATIONS RESEARCH
UNIVERSITY OF NORTH CAROLINA AT CHAPEL HILL
CHAPEL HILL, NORTH CAROLINA 27599
USA
E-MAIL: haipeng@email.unc.edu

DEPARTMENT OF STATISTICS
TEXAS A&M UNIVERSITY
COLLEGE STATION, TEXAS 77843
USA
E-MAIL: jianhua@stat.tamu.edu